\documentclass[pre,floatfix,superscriptaddress,twocolumn,9pt]{revtex4-1}

\usepackage{amsmath}
\usepackage{amssymb}
\usepackage{amsfonts}
\usepackage{graphicx}
\usepackage{color}
\usepackage{bm}
\usepackage[english]{babel}
\usepackage{hyperref}
\usepackage{siunitx}
\usepackage{subfigure}

\newcommand{\figurewidth}{.45\textwidth}

\newcommand{\state}[1]{$\mathcal{S}_{#1}$}

\begin{document}

\title{High-resolution Markov state models for the dynamics of Trp-cage miniprotein constructed over slow folding modes identified by state-free reversible VAMPnets}

\author{Hythem Sidky}
\affiliation{%
  Pritzker School of Molecular Engineering, %
  University of Chicago, %
  Chicago, Illinois 60637%
}

\author{Wei Chen}
\affiliation{
	Department of Physics, %
	University of Illinois at Urbana-Champaign, %
	1110 West Green Street, Urbana, Illinois 61801
}

\author{Andrew L. Ferguson}
\email{Author to whom correspondence should be addressed: \mbox{andrewferguson@uchicago.edu}}
\affiliation{%
  Pritzker School of Molecular Engineering, %
  University of Chicago, %
  Chicago, Illinois 60637%
}

\begin{abstract}
\noindent State-free reversible VAMPnets (SRVs) are a neural network-based framework capable of learning the leading eigenfunctions of 
the transfer operator of a dynamical system from trajectory data. In molecular dynamics simulations, these data-driven collective variables (CVs) capture 
the slowest modes of the dynamics and are useful for enhanced sampling and free energy estimation. In this work, we employ 
SRV coordinates as a feature set for Markov state model (MSM) construction. Compared to the current state of the art, MSMs constructed from SRV coordinates are more robust to the choice of input features, exhibit faster implied timescale convergence, and permit the use of shorter lagtimes to construct higher kinetic resolution models. 
We apply this methodology to study the folding kinetics and conformational landscape of the Trp-cage miniprotein. Folding and unfolding mean first passage times are in good agreement 
with prior literature, and a nine macrostate model is presented. 
The unfolded ensemble comprises a central kinetic hub with interconversions to 
several metastable unfolded conformations and which serves as the gateway to the folded ensemble. The folded ensemble comprises the native state, a partially unfolded intermediate ``loop'' state, and a previously unreported short-lived intermediate that we were able to resolve due to the high time-resolution of the SRV-MSM. We propose SRVs as an excellent candidate for 
integration into modern MSM construction pipelines. 
\end{abstract}

\maketitle

\section{\label{sec:intro}Introduction}

Molecular dynamics (MD) simulations are an indispensable tool in the study of the conformational, thermodynamic, and kinetic properties of biomolecular systems. Advances in MD software and hardware have enabled access to millisecond timescales at atomistic
resolution, but a major challenge is how to best analyze these large 
simulated trajectories to extract experimentally-meaningful kinetic and thermodynamic quantities. 

Markov State Models (MSMs) have emerged as a powerful framework for analyzing MD simulations and recovering 
dynamical properties of interest.~\cite{Husic2018} Their primary innovation is to discretize high-dimensional molecular conformational space into coarse-grained states, wherein the dynamical interconversions between microstates within a macrostate are fast relative to transitions between macrostates. Accordingly, the macrostate dynamical transitions are approximately memoryless (i.e., Markovian) and can be modeled by a master equation.~\cite{Pande2010} Protein folding has benefited immensely from developments in MSM methodology which 
have pushed the limits of recoverable long-term kinetics while simultaneously yielding insight into microscopic
quantities.~\cite{Prinz2011,Plattner2017} Nevertheless, the quality of 
a MSM is highly dependent on the input features, state space decomposition, and a number of parameters chosen during 
its construction. This has motivated research into optimizing each stage of the MSM pipeline including 
theory,~\cite{Prinz2011a, Noe2013} basis selection,~\cite{Schwantes2013, Noe2015}, clustering,~\cite{Husic2017}
and validation.~\cite{McGibbon2015, Husic2016}

The current state of the art in MSM construction involves the use of time-lagged independent component analysis 
(TICA)~\cite{Pande2010, Schwantes2013, Perez-Hernandez2013} to identify a linearly-optimal combination of input features which
maximizes their kinetic variance. Clustering is then performed in this slow subspace to produce the states between which interconversion rates are estimated. TICA has all but superseded structural clustering based on metrics such as minimum 
root mean square distance (RMSD) that tend to capture motions of high structural variance as opposed to the desired 
slowest motions.~\cite{Husic2018, Perez-Hernandez2013} A recently proposed alternative to MSMs are VAMPnets, an 
artificial neural network (ANN) approach that seeks to replace the entire MSM pipeline.~\cite{Mardt2018} VAMPnets are a very promising new technique, but as an end-to-end replacement to MSM construction cannot yet be interfaced with the extensive machinery and extensions developed for MSMs such as statistical error estimators, rare event sampling techniques, and incorporation of experimental constraints.~\cite{Mardt2018}

In a recent work, we proposed state-free reversible VAMPnets (SRVs)~\cite{Chen2019} as a deep learning framework based on VAMPnets~\cite{Mardt2018}, which themselves are based on deep canonical correlation analysis (DCCA)~\cite{Andrew2013}. Contrary to VAMPnets, SRVs were  designed not to approximate MSMs but rather to directly learn nonlinear approximations to the slowest dynamical modes of a molecular system obeying detailed balance. The approach is founded on the variational approach to conformational dynamics (VAC), which defines a variational principle for the slowest eigenfunctions of the transfer operator that propagates state functions through time.~\cite{noe2013variational, nuske2014slow} The essence of our approach is to use twin-lobed neural networks to learn the best nonlinear basis set to pass to the linear variational problem defined by the VAC. The VAC then furnishes the optimal eigenvector approximations of the transfer operator ordered by decreasing implied timescales. Following VAMPnets, we deviate from DCCA in choosing as our loss function the VAMP-2 score informed by the variational approach to Markov processes (VAMP) principle~\cite{Mardt2018}. Contrary to VAMPnets, we modify our network architecture to directly approximate 
the slow modes of the transfer operator rather than soft metastable state assignments, and employ the 
variational approach under detailed balance to approximate the slow modes of equilibrium dynamics.
(Our prefix ``state-free reversible'' reflects these two key differences.) SRVs can also be viewed as a
multi-dimensional generalization of variational dynamics encoder~\cite{Hernandez2017}, a variational analog to
time-lagged autoencoders~\cite{Wehmeyer2018}, and are closely related to kernel
TICA.~\cite{Schwantes2015} 

In this work, we demonstrate the utility of employing the slow modes recovered by SRVs as a basis within which to 
construct MSMs. This study was motivated by the hypothesis that compared to MSMs based on linear TICA approximations to the transfer operator eigenfunctions, MSMs constructed from the nonlinear SRV approximations would permit the use of shorter lagtimes and therefore furnish models with  higher kinetic resolution. Whereas VAMPnets perform nonlinear featurization, slow-mode estimation, and soft clustering into metastable states macrostates,~\cite{Wehmeyer2018} SRVs perform only the 
first two steps. The final step of MSM construction is performed using standard protocols utilizing the slow modes learned by SRVs rather than TICA coordinates. In this manner, we take advantage of the large body of theoretical work and mature numerical implementations developed for MSM construction~\cite{Husic2018, Pande2010, schererpyemma2015, Beauchamp2011}
where SRVs serve as a modular replacement for TICA. SRV-MSMs are shown to perform better than TICA-MSMs under cross validation, offer more flexibility and robustness in feature selection, and converge implied
timescales quicker, allowing for shorter lagtimes and ultimately a higher resolution kinetic model. VAMPnets and SRV-MSMs perform comparably, but, as we will show, 
the SRV-MSM exhibits slightly faster convergence of the implied timescales and enables access to the statistical error estimators,~\cite{Shirts2008} multi-ensemble approaches,~\cite{Wu2016} and other extensions developed for MSMs.~\cite{Prinz2011b, Mardt2018}

We demonstrate SRV-MSMs in an application to an ultra-long 208 $\mu s$ explicit solvent simulation of the K8A mutant of Trp-cage TC10b at 290 K performed by D.E.~Shaw Research.~\cite{Lindorff-Larsen2011} Trp-cage is a fast-folding miniprotein that has been the 
subject of numerous experimental~\cite{Barua2008, Meuzelaar2013} and 
computational studies.~\cite{Marinelli2009, Zhou2001, Zhou2003, Juraszek2006, Meuzelaar2013, Kim2015} Despite its status as an archetypal miniprotein for the testing of new computational methods, its kinetic behavior remains incompletely understood. Given the sensitivity of the 
Trp-cage folding landscape to mutations~\cite{Barua2008} and termini,~\cite{English2015} a direct comparison of the behavior 
of different mutants is not possible. The K8A mutant of Trp-cage TC10b considered in this work has been previously studied by Dickson \& Brooks~\cite{Dickson2013}, who determined that the Trp-cage unfolded ensemble 
displays two-state behavior. Su\'{a}rez et al.~\cite{Suarez2016}\ analyzed the same data using non-Markovian techniques to determine 
mean first passage times (MFPT) between the folded and unfolded states.
Deng et al.\ conducted perhaps the most comprehensive study of the the kinetics of this data to date,~\cite{Deng2013, Levy2013} identifying two 
representative folding mechanisms: the hydrophobic collapse of Trp-cage into a molten globule followed by the formation of 
the N-terminal $\alpha$-helix and native core (nucleation-condensation), and the pre-formation of the $\alpha$-helix in an extended unfolded 
state then the joint formation the $3_{10}$ helix and hydrophobic core (diffusion collision). 
The diffusion-collision mechanism is identified as the dominant folding pathway with a substantially smaller transit time of 3 ns, compared to 42 ns for nucleation-condensation.
The high kinetic resolution of the model furnished by SRV-MSMs in the present work establishes new understanding of the Trp-cage folding mechanism, and demonstrates SRVs as a valuable tool in the construction of high kinetic resolution MSMs.


\section{\label{sec:methods}State-free reversible VAMPnet MSM construction} \label{sec:MSM}

We now proceed to describe our SRV-MSM construction pipeline, comprising the following six steps: (i) feature selection, (ii) SRV learning of the slow modes, (iii) definition of microstates and microstate transition rates by k-means clustering in the SRV coordinates, (v) definition of MSM macrostates and macrostate transition rates by spectral clustering of the microstate transition matrix, and (vi) comparison of the resulting SRV-MSM with a TICA-MSM and VAMPnets. The molecular simulation we study  is a 208 $\mu s$ explicit solvent K8A mutant of Trp-cage TC10b simulation performed by D.E. Shaw Research using the CHARMM22* forcefield and containing approximately $10^6$ snapshots at intervals of 200 picoseconds.~\cite{Lindorff-Larsen2011} 


\subsection{Molecular feature selection}

\begin{figure}[h!]
	\begin{center}
        \includegraphics[width=\figurewidth]{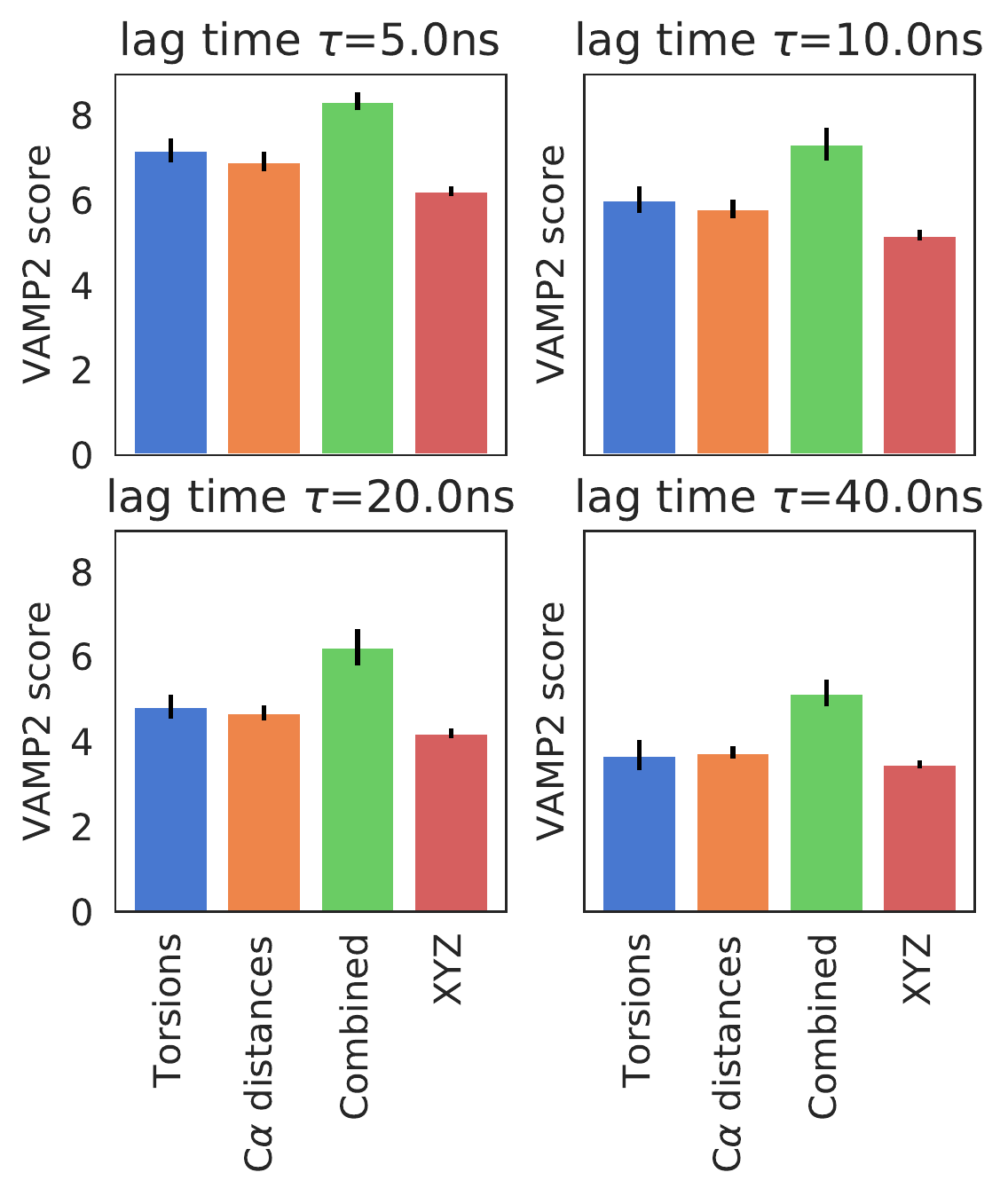}
        \caption{Molecular feature selection. 
        VAMP-2 scores of the five slowest processes for various feature transformations of the Trp-Cage trajectory calculated at a variety of lagtimes $\tau$: backbone and sidechain torsions (torsions), C$\alpha$ pairwise atom distances (C$\alpha$ distances), 
a combination of the previous two features (combined), and the  aligned Cartesian coordinates of the entire molecule (XYZ).
        The combined featurization comprising backbone and sidechain torsions and $C_{\alpha}$ pairwise 
	    distances is superior across all tested lagtimes and is used for all subsequent analysis. 
        } 
        \label{fig:feature_vamp2}
	\end{center}
\end{figure}

In order to perform slow variable discovery we must first define the set of features derived from each instantaneous configuration of the molecular system that will be used to represent the trajectory to  the learning algorithm. Scherer et al.~\cite{Scherer2018} have
recently shown that feature choices can be optimized directly through a variational principle based on VAMP scoring without requiring construction of the entire kinetic model. The scoring method, known as VAMP-2 scoring,~\cite{Wu2017} is the sum of the squared estimated eigenvalues
of the transfer operator. Under this variational approach, larger cross-validated VAMP-2 scores correspond to more kinetically accurate models and the cross-validated test score is bounded from above by the true kinetic model. 
We employ this method of variational feature selection using backbone and sidechain torsions, C$\alpha$ pairwise atom distances, 
a combination of these two features, and the aligned Cartesian coordinates of the entire molecule.
Figure~\ref{fig:feature_vamp2} shows the result of ten-fold cross-validated VAMP-2 scoring for the aforementioned feature sets at different lagtimes $\tau$ 
using the top ten eigenvalues. It is clear the the combined set of torsions and C$\alpha$ pairwise distances contain more 
kinetic variance at  all lagtimes considered,~\cite{Noe2015} and hence should be preferred over the other feature sets. The aligned Cartesian coordinates 
consistently underperform the other choices. We use the combined set of torsions and C$\alpha$ pairwise distances for all further analysis unless 
otherwise stated.

\subsection{SRVs outperform TICA-MSMs under cross-validation} \label{SRVs}

TICA-MSM models were built using PyEMMA~\cite{schererpyemma2015} 2.5.4 following the general prescriptions 
outlined in Ref.~\cite{Wehmeyer2018Introduction}. Using the combined feature set, the number of TICA dimensions, 
TICA lagtime, and number of cluster centers were optimized under VAMP-2 scoring. It was determined 
that five TICs were optimal and that the VAMP-2 score was insensitive beyond a sufficient number of cluster centers over the range 5 to 500, within which 200 was chosen.
SRVs were trained using the SRV package~(\url{https://github.com/hsidky/srv}) with the default architecture of two hidden layers with 
100 neurons each, and $\tanh$ activation functions. We specified a batch size of 500,000, a learning rate of $0.01$, and employed batch
normalization within all hidden layers. No early stopping or weight decay was used to maximize data utilization and 
avoid having to tune regularization strength. Instead, we screened for number of training epochs as part of the VAMP-2 
optimization. All code used for model screening, selection, and generation of results can be found in the repository \url{https://github.com/hsidky/srv-trpcage}.

\begin{figure*}
	\begin{center}
        \includegraphics[width=\textwidth]{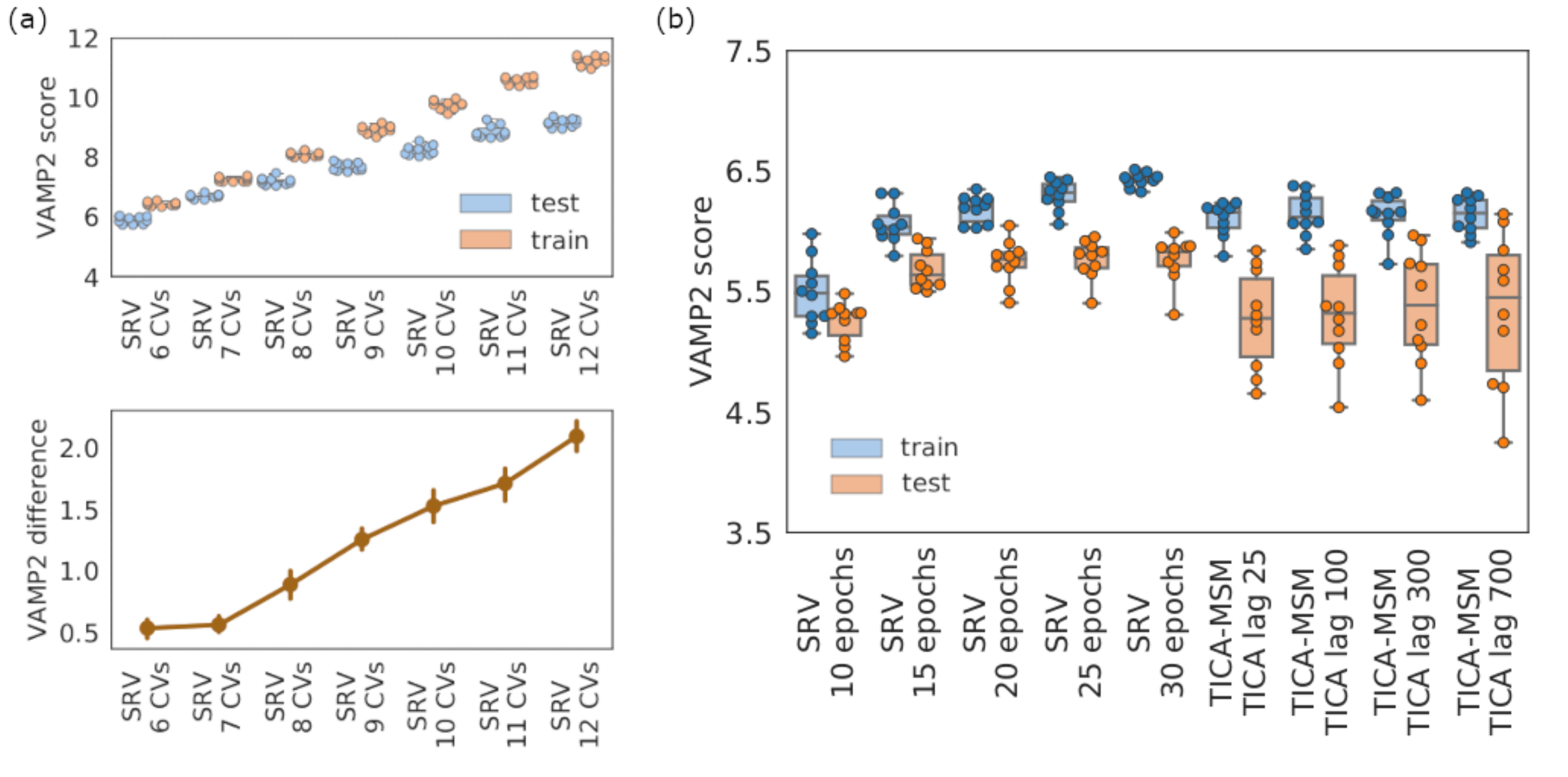}
        \caption{SRV and TICA-MSM model validation. (a) Ten-fold cross-validated VAMP-2 scores for SRV models containing an increasing number of SRV coordinates constructed at a lagtime of $20$ ns (upper panel). An increase in the gap between testing and training scores (lower panel) indicates the onset of overfitting and motivates the selection of a seven SRV coordinate model as that best supported by the data. (b) Cross validation of the SRV training epochs and TICA lagtime in steps (5 steps = 1 ns) hyperparameters against the VAMP-2 score demonstrate SRVs have higher train and test scores and narrower distributions, which is indicative of model robustness and generalizability.}
	\label{fig:hde_vamp2}
        \end{center}
\end{figure*}

We used ten-fold cross-validated VAMP-2 scores to compare the quality of different SRV models and TICA-MSM models. 
Specifically, to maximize the similarity between the train and test data distributions, we first divided 
the full 208 $\mu s$ trajectory into 100 equal segments which are treated as independent trajectories for the  purposes of our comparative analysis. The segments are then shuffled and 
subsampled as part of train-test split procedure for each fold. This approach has the drawback of losing transitions 
across the individual segments, but it ensures that the conformational distribution over the complete trajectory is well represented in both the training and testing sets. Note that here we choose to compare the VAMP-2 scores of the TICA-MSMs directly to
the SRVs rather than a subsequent SRV-MSM. The primary reason for this is that we want to make clear 
the contribution of the SRV coordinates themselves to the kinetic content without additional processing. We present a comparison between the TICA-MSM and SRV-MSM implied timescales later on in Section \ref{impliedTimescales}.

To determine the number of eigenvalues to retain for cross validation, we calculate train and test VAMP-2 
scores for SRVs of increasing dimensionality. From Figure~\ref{fig:hde_vamp2}a, there is a marked increase in the gap 
between training and testing scores after seven dimensions, which is indicative of overfitting and motivating our choice of the seven SRV eigenvector model as that best supported by the data. The SRVs were trained at a lagtime of 20 ns which is the same lagtime used for TICA-MSM construction and VAMP-2 scoring.

Figure~\ref{fig:hde_vamp2}b presents the cross-validated VAMP-2 scoring for the SRV and TICA-MSM models. 
Both classes of models perform well but with some significant differences. While the distributions of training scores are very similar, SRVs display remarkable 
consistency in test scores compared to the TICA-MSMs. TICA-MSM test scores vary considerably between folds, which is 
indicative of model sensitivity to training data and characteristic of overfitting. The SRVs show consistent improvement in training scores with the number of training epochs, but the plateau in the  testing score and widening gap between the training and test scores after 20 epochs signals overfitting. The 30 epoch model still yields a marginally higher test VAMP-2
score than other epochs, which is our selection, but the difference between 20, 25, and 30 epochs is insignificant. The TICA lagtime does not appear to have much of an impact on train or test score means, although we  do note a marked increase in the testing variance for the largest lagtime of 700 steps (140 ns). 

\begin{figure}
	\begin{center}
        \includegraphics[width=\figurewidth]{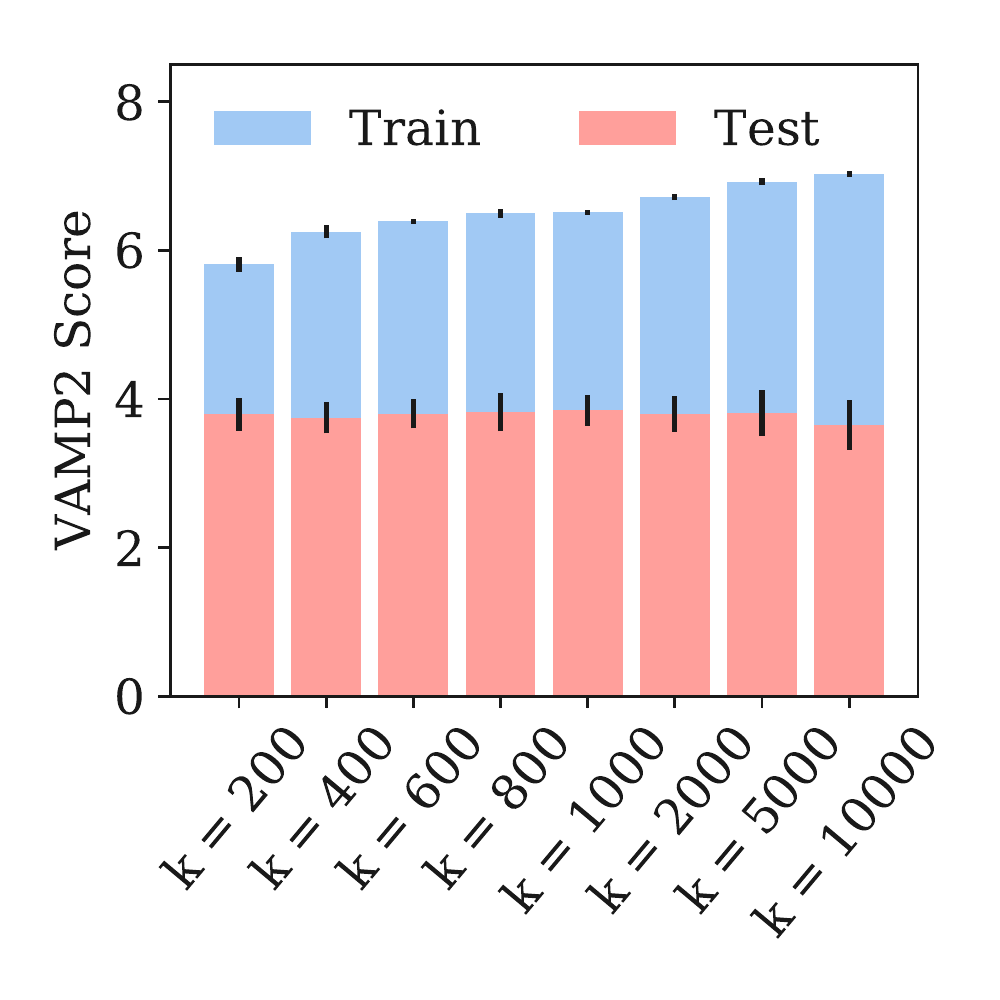}
        \caption{
            Ten-fold cross validated VAMP-2 scores for RMSD-MSMs. Although the training score increases with number of states, the widening gap between the test and training scores is indicative of overfitting.
        } 
        \label{fig:vamp_scoring_rmsd}
	\end{center}
\end{figure}

For comparison, Figure~\ref{fig:vamp_scoring_rmsd} shows the result of ten-fold cross validation for RMSD-based MSMs for 
increasing number of microstates $k$. An RMSD-MSM with $k$ = 25,000 microstates was previously utilized by Deng et
            al.~\cite{Deng2013}\ in the analysis of the D.E.\ Shaw 208 $\mu s$ Trp-cage simulation considered herein. The training VAMP-2 score increases with the number of microstates, which 
results higher implied timescales and seemingly better performance. However, the test scores remain approximately constant, with a small decrease at $k$ = 10,000. 
This widening gap between testing and training scores is indicative of overfitting, and although the RMSD-MSM training scores are similar to TICA-MSM and SRVs, the test scores are significantly worse for all values of $k$.

The higher 
train and test scores of the SRVs and improved variance over TICA-MSMs and RMSD-MSMs indicate that they are more kinetically 
accurate, capture more information about the system dynamics, and thus present an excellent basis in which to construct kinetic models of the system dynamics.

\subsection{SRVs are robust to the choice of feature set} \label{features}

We showed in Section \ref{SRVs} that using the same optimized feature set, SRVs outperform TICA-MSMs under cross-validation. We 
now address the situation where sub-optimal features are used to construct both models. Empirical evidence suggests that 
it may be useful~\cite{Scherer2018} to generate a bank of distance or contact-based features that are nonlinear featurizations of the atomic coordinates to improve MSM quality. Examples of these transformations include reciprocals, logarithms, polynomials, or
exponentials of pairwise distances. Improvement is possible since TICA is restricted to discover linear combinations of the input features, and nonlinear feature engineering can introduce nonlinearities into the model. Since SRVs are based on a deep learning architecture, the universal approximation theorem~\cite{hassoun1995fundamentals, chen1995universal} asserts that they should, by employing sufficiently many hidden nodes, be capable of discovering nonlinear feature transformations to maximize the kinetic variance from rather poor choices of input feature sets without extensive feature engineering. Here, we test this conjecture by omitting the  backbone and sidechain torsions from the feature set.

\begin{figure*}[ht!]
	\begin{center}
        \includegraphics[width=\textwidth]{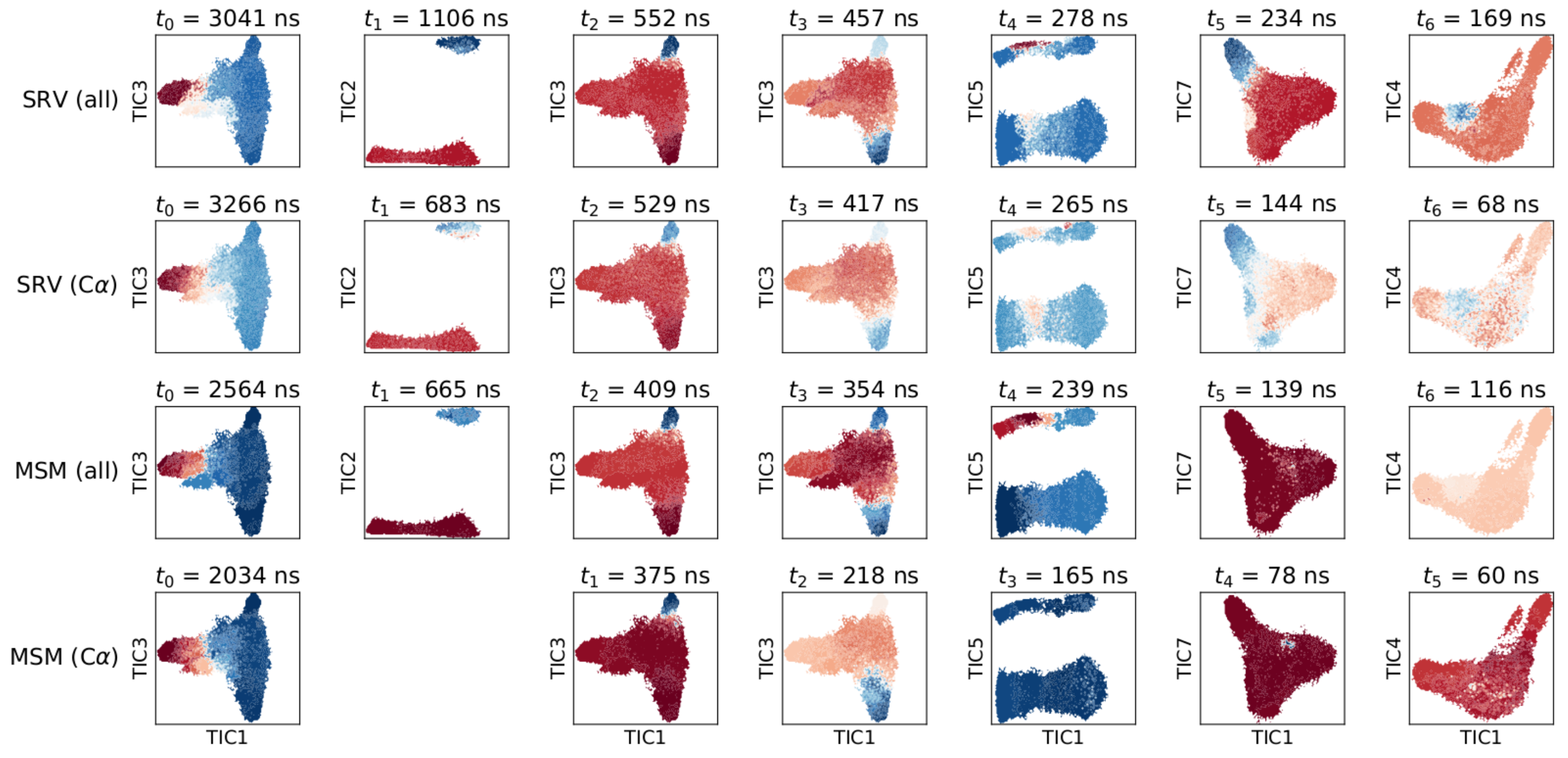}
        \caption{Visualization of the top seven eigenvectors of the SRV and TICA-MSM computed under two different feature sets: $C\alpha$ pairwise distances only ($C\alpha$), and $C\alpha$ pairwise distances plus backbone and sidechain torsions (all). The eigenvectors are projected onto the leading TICA coordinates TIC1-7 obtained in construction of the TICA-MSM under the $C\alpha$ pairwise distances plus backbone and sidechain torsions feature set. The implied timescales associated with each eigenvector is printed above each panel. SRVs (first row) and TICA-MSMs (third row) trained on $C\alpha$ pairwise distances and backbone and sidechain torsions discover the same leading five slowest modes, although the SRV discovers slower timescales. Excluding torsions from the input feature set renders the second leading mode TIC2 (second column, $t_1 \approx$ 1 $\mu$s) invisible to the TICA-MSM (fourth row), whereas a SRV (second row) can adequately resolve it and the remaining higher-order modes by forming nonlinear combinations of the $C\alpha$ pairwise distances input features.
        } 
        \label{fig:hde_msm_feat_comp}
	\end{center}
\end{figure*}

Figure~\ref{fig:hde_msm_feat_comp} presents a visualization of the top seven SRV and top seven TICA-MSM eigenvectors constructed over two feature sets: one comprising $C\alpha$ pairwise distances only, and one comprising $C\alpha$ pairwise distances plus backbone and sidechain torsions. The eigenvectors are projected onto TICA coordinates (TIC1-7)
obtained in construction of the  TICA-MSM under the $C\alpha$ pairwise distances plus backbone and sidechain torsions feature set. We choose to visualize along TICA coordinates since they contain more variance than the 
SRV or TICA-MSM eigenfunctions, which makes them more suitable for visualization purposes. The key difference between the feature sets 
emerges in the second slow mode (TIC2, second column) learned from the combined  $C\alpha$ pairwise distances plus backbone and sidechain torsions, where the SRV constructed using only $C\alpha$ pairwise distances (second row) 
is able to learn a transition along TIC2 whereas the MSM trained on only $C\alpha$ pairwise distances data (fourth row) fails to do so. Furthermore, 
the SRV trained only on $C\alpha$ pairwise distances (second row) successfully discovers the remaining higher-order modes with only a minor degradation in the implied timescales relative to the SRV trained on torsions and $C\alpha$ pairwise distances (first row). The dynamical motion associated with TIC2 has a timescale 
of $t_1 \approx 1$ $\mu$s, and by failing to account for it a significant contribution to the kinetic variance is lost. The nonlinear nature of the SRV enabled it to form nonlinear combinations of the $C\alpha$ pairwise distances input features to discover the dynamical motions associated with torsional angles necessary to resolve this mode. SRVs are therefore able to discover an important slow dynamical mode that is invisible to a TICA-MSM presented with the same data. This capability is particularly valuable in extracting maximal kinetic variance from suboptimal input feature sets, and can be used in concert with VAMP scoring to identify the optimal feature set without extensive manual feature engineering.


\subsection{Implied timescales of SRV-MSMs exhibit faster convergence than TICA-MSMs and VAMPnets} \label{impliedTimescales}

We demonstrated in Section \ref{SRVs} that the leading SRV eigenvectors present a good basis in which to represent the long time system dynamics, and that cross-validation with respect to the VAMP-2 score showed the kinetic model based on the top seven SRV eigenvectors to be best supported by the data. We now proceed to use these coordinates to construct a SRV-MSM with which we may propagate the long-time evolution of the system and analyze for its macrostate configurational discretization, stationary state occupancy probabilities, dwell times, and transition rates.

\begin{figure}
	\begin{center}
        \includegraphics[width=\figurewidth]{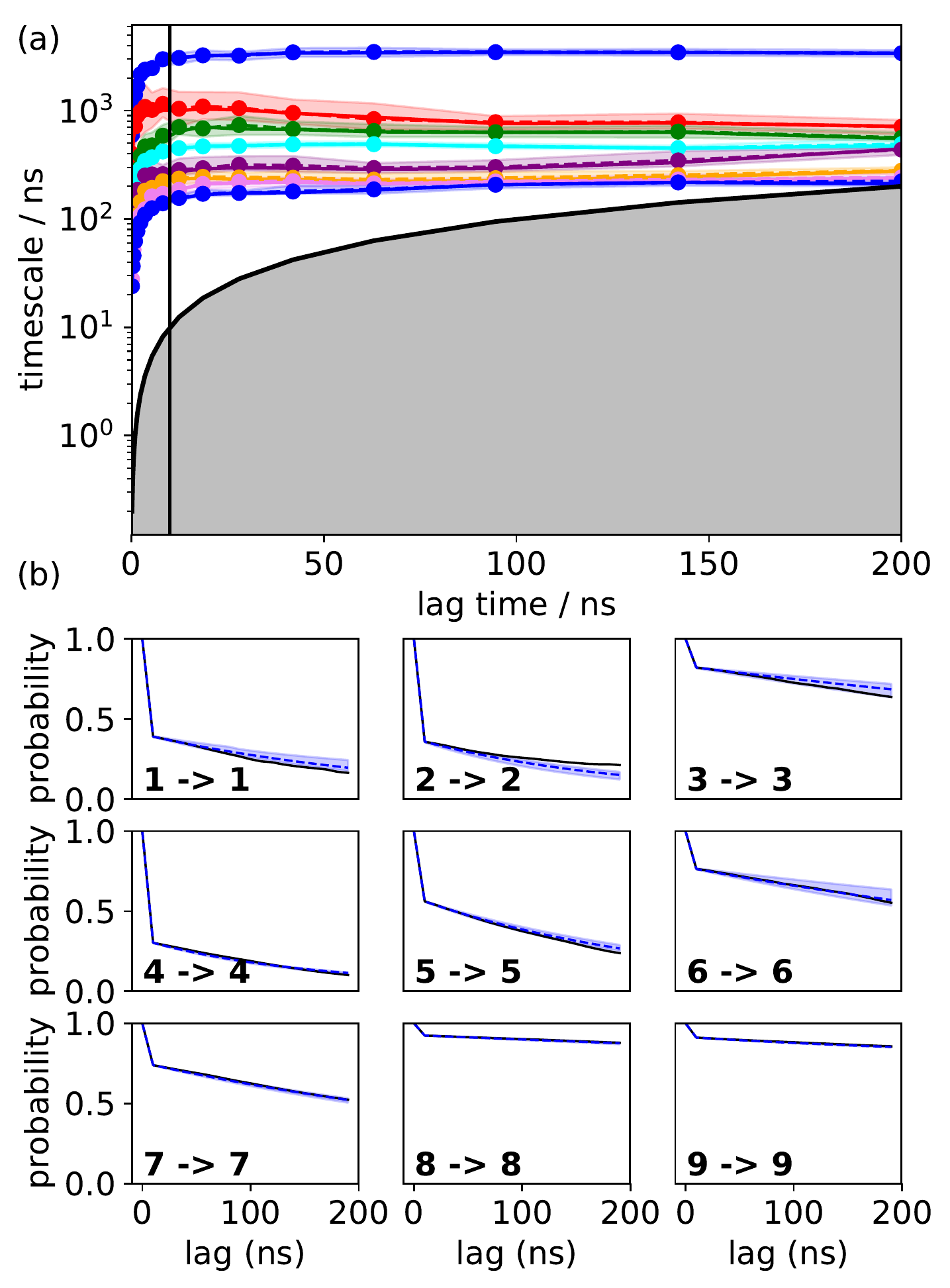}
        \caption{Validation of the SRV-MSM. (a) Convergence of the eight implied timescales of the nine-macrostate SRV-MSM as a function of lagtime. Solid lines indicate maximum likelihood result while dashed lines show the Bayesian ensemble means. The SRV-MSM timescales converge at a lagtime of $\tau$ = 10 ns (vertical line). The black solid curve marks equality of  the implied timescale and lagtime and delimit the shaded region where the implied timescales are shorter than the lagtime and cannot be resolved. (b) The Chapman-Kolmogorov (CK) test comparing the probabilities of remaining within each of the nine macrostates as a function of lagtime predicted by a SRV-MSM constructed at a $\tau$ = 10 ns lagtime (dashed blue line) and those computed from a SRV-MSM constructed at the particular lagtime (solid black line). In both panels the shaded areas represent 95\% confidence intervals. Rapid convergence of  the implied timescales and agreement of the predicted and computed transition probabilities confirm the dynamic validity of the SRV-MSM.
        } 
        \label{fig:msm_its_cktest}
	\end{center}
\end{figure}

The SRV-MSM was constructed using the PyEMMA software package~\cite{schererpyemma2015}. A microstate transition matrix comprising 100 microstates, where this number was selected by hyperparameter optimization, was constructed by performing k-means clustering of projections of the simulation trajectory into the leading seven SRV eigenvectors. Diagonalization of the microstate transition matrix reveals eight leading timescales followed by a spectral gap, motivating the construction of a nine macrostate SRV-MSM. Figure~\ref{fig:msm_its_cktest}a
shows the eight implied timescales to converge extremely rapidly with lagtime $\tau$, enabling selection of a very short $\tau$ = 10 ns lagtime and construction of a high temporal resolution SRV-MSM. To validate the resulting SRV-MSM, we conduct a Chapman-Kolmogorov (CK) test. The CK test compares the transition probabilities between pairs of states $i \rightarrow j$ at a lagtime of $k \tau$ predicted by a model constructed at a lagtime $\tau$ and that computed directly from a model constructed at a lagtime $k \tau$. Figure~\ref{fig:msm_its_cktest}b presents the results of the CK test for within state (i.e., $i \rightarrow i$) transitions. We observe that our $\tau$ = 10 ns lagtime model performs excellently in predicting the transition probabilities even out to very long times of $k \tau$ = 200 ns. This CK analysis demonstrates that the SRV-MSM kinetic model is Markovian for a $\tau$ = 10 ns lagtime, where this high temporal resolution is made possible by our use of SRV eigenvectors for microstate clustering. 


\begin{figure}
	\begin{center}
        \includegraphics[width=\figurewidth]{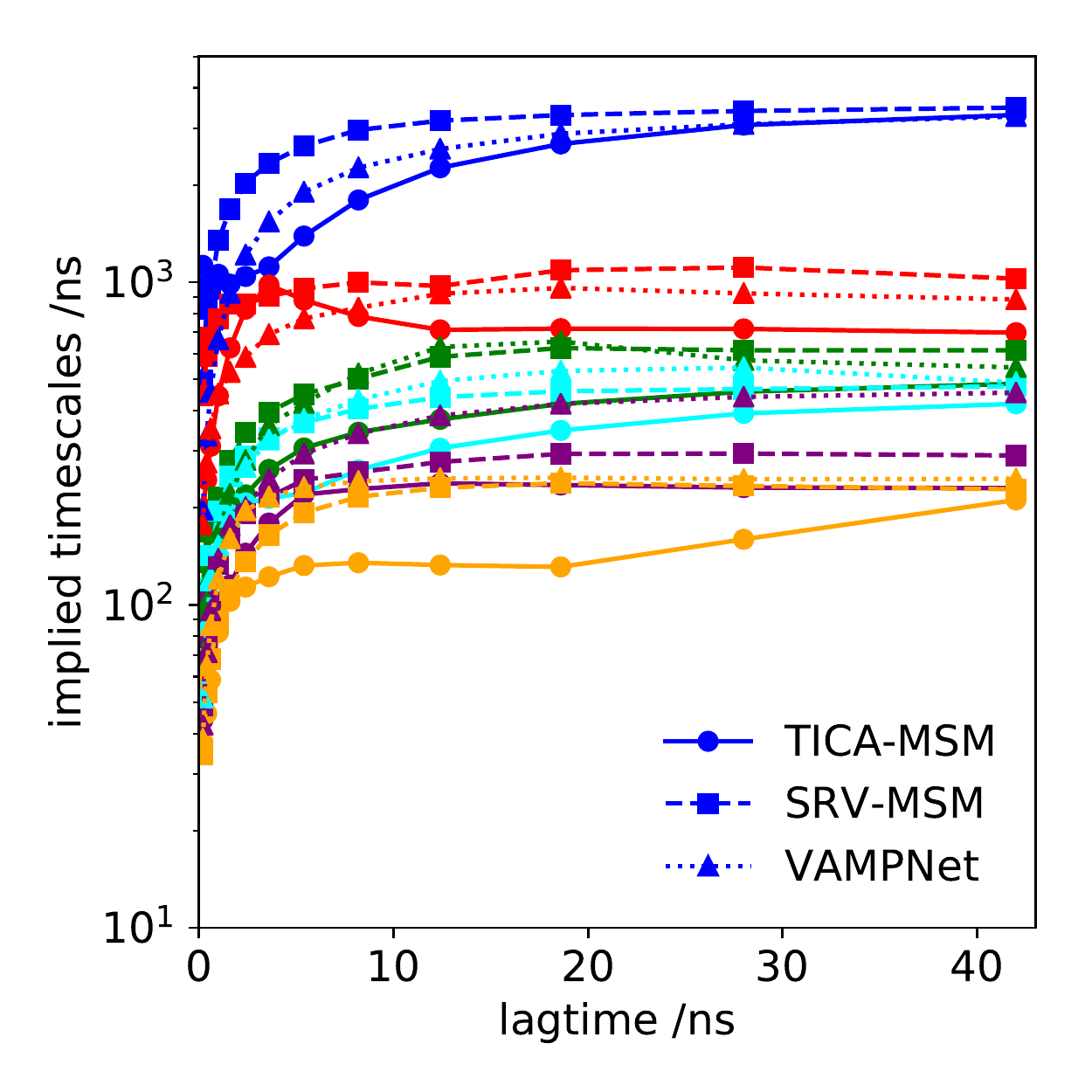}
        \caption{Close-up of the convergence of the leading six implied timescales as a function of lagtime for TICA-MSM (solid line, circles), SRV-MSM (dashed line, squares), and VAMPnets (dotted line, triangles). The SRV-MSM converges the implied timescales at approximately five times shorter lagtimes than VAMPnets or TICA-MSMs, enabling the construction of an extremely high time resolution MSM.} 
        \label{fig:tica_hde_its_comparison}
	\end{center}
\end{figure}

To compare SRV-MSMs, TICA-MSMs, and VAMPnets, we present in Figure~\ref{fig:tica_hde_its_comparison} a close-up of the convergence of the implied timescales as a function of lagtime for the optimized TICA-MSM (Section \ref{SRVs}), SRV-MSM, and an equivalent nine-state VAMPnet constructed at the same $\tau$ = 10 ns lagtime. Due to the congested nature of this plot, we choose to plot only the leading six implied timescales for clarity. The SRV-MSM converges the slowest implied timescale at approximately five times shorter lagtimes than the TICA-MSM or VAMPnets. Convergence of the higher-order timescales is similar for VAMPnets and the SRV-MSM, whereas the TICA-MSM fails to converge to the same values even at quite long lagtimes. This trend can be attributed to the fact that the SRV-MSM and VAMPnets are able to learn nonlinear transformations of the input coordinates and therefore better resolve slower processes that are invisible to the inherently linear TICA-MSM (cf.~Section \ref{features}).

In summary, the convergence of the implied timescales and validation of the CK test demonstrates that the SRV-MSM based on seven SRV coordinates (selected by cross-validating the training and testing VAMP-2 scores), nine metastable macrostates (selected by a gap in the microstate eigenvalue spectrum after the eighth non-trivial eigenvalue), and a lagtime of $\tau$ =  10 ns (estimated by convergence of implied timescales) presents a good kinetic model for the long-term system dynamics at a higher temporal resolution than is accessible using a TICA-MSM. This demonstrates the value of a modular replacement of TICA coordinates conventionally used for microstate clustering by SRV coordinates within an MSM pipeline in order to achieve higher temporal resolution MSM models while preserving access to the large body of tools and infrastructure developed for the construction, validation, and analysis of Markov state models.~\cite{Bowman2010, Wu2016, Shirts2008, Prinz2011b}

\section{\label{sec:results}Trp-Cage Analysis}

\begin{figure*}[t]
	\begin{center}
        \includegraphics[width=\textwidth]{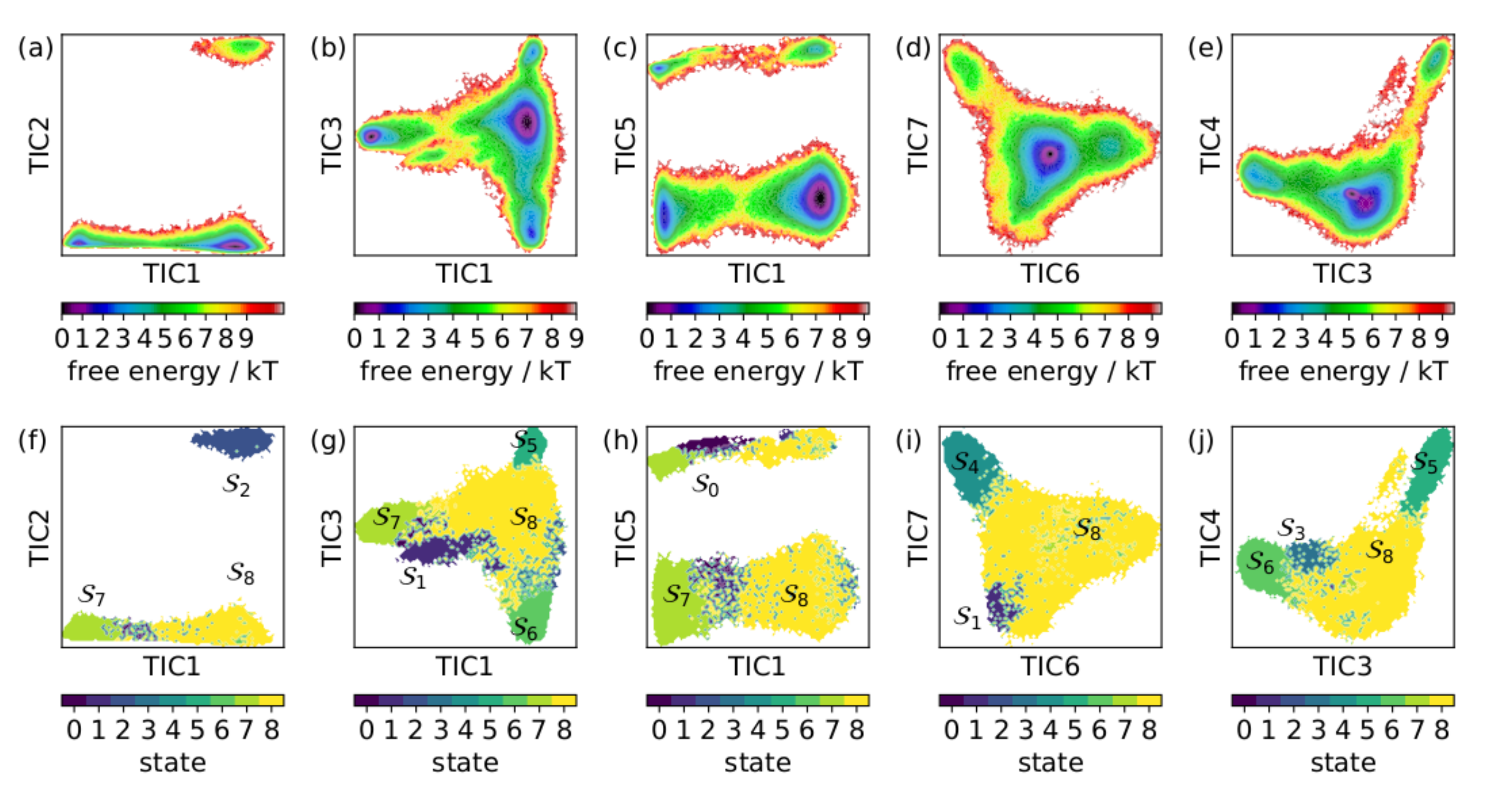}
        \caption{Free energy surfaces and macrostate clustering visualizations projected into the leading TICA coordinates (TIC1-7) for visualization purposes. (a-e) Free energy surface projected onto various pairwise combinations of TICs. (f-j) Metastable macrostate assignments \state{0-8} computed by PCCA\texttt{++} within the same TIC projections. Macrostate \state{7} contains the folded state.
        }
        \label{fig:hde_msm_fes}
	\end{center}
\end{figure*}

We now commence our analysis of Trp-cage folding dynamics based on the SRV-MSM constructed and validated in Section \ref{sec:MSM}. It is first useful to visualize low-dimensional free energy landscapes illustrating the nine macrostates in order to generate an overview of the relative locations of the metastable macrostates of the model. As is customary,~\cite{schererpyemma2015} we visualize the free energy landscapes in the leading TICA coordinates as good high-variance collective variables in which to construct and display the free energy surface. We emphasize that these TICA coordinates are used exclusively as convenient linear collective variables that support good visualizations, whereas the MSM is constructed from the nonlinear SRV coordinates. To obtain more accurate free energy estimates along the TICA coordinates, we reweight each frame of the simulation trajectory by the associated values of the stationary distribution computed from the 100 microstate transition matrix, project these weighted data onto the leading TICA coordinates TIC1-7, and then estimate free energy surfaces from the empirical probability distributions within this space. We display selected 2D projections of the free energy surface within pairs of TICs in the top row of Figure~\ref{fig:hde_msm_fes}, and in the bottom row show the clustering into the nine metastable macrostates \state{0-8} computed from PCCA\texttt{++} spectral clustering.~\cite{Roblitz2013, Deuflhard2005, Kube2007}

We caution against over-interpreting low-dimensional free energy landscape projections, but the gross features of the landscape are a folded state represented by  \state{7} connected to the large unfolded ensemble of states \state{2-6,8} by a narrow neck. \state{2-6} represent structured metastable conformations within the unfolded ensemble. \state{2} corresponds to an extended conformation with outwardly rotated prolines, \state{3} a crossed conformation with a minor central hairpin, \state{4} a braided hairpin-like structure, \state{5} a hairpin, and \state{6} a configuration with a collapsed N-terminus and an extended C-terminus. We provide a finer-grained molecular-level description and visualization of the states when we discuss the macrostate transition matrix. 

The first non-trivial right eigenvector of the macrostate transition matrix describes the transition to and from \state{7} which represents  
the folded state. Based on this definition, there are 12 observed folding and unfolding events in the trajectory, 
which agrees with the value reported by Lindorff-Larsen et al.~\cite{Lindorff-Larsen2011}, who used a native 
contacts-based definition of folded and unfolded states. The fraction of native contacts, $Q$, has 
been previously shown to accurately characterize the thermodynamics of protein folding in and out of the native state.~\cite{Best2013, Meshkin2017} Conversely, this figure is in poor agreement with the 31 folding transitions reported by Deng et al.~\cite{Deng2013} who use an 
RMSD-based definition of folding. This choice of an RMSD distance introduces a number of additional rapid folding transitions, and -- based on the good agreement between the $Q$-based and MSM-based definitions of folding -- suggests that this structural measure is a poor proxy for kinetic proximity.

We report in Table~\ref{table:mfpt} the mean first passage times (MFPTs) into and out of the the folded state, \state{7}, with uncertainties estimated using a Bayesian scheme emnploying 50 samples~\cite{Wehmeyer2018Introduction, Noe2008, Trendelkamp-Schroer2015}. Our calculated MFPTs are in good agreement with Lindorff-Larsen et
al.~\cite{Lindorff-Larsen2011} who report values of 14.4 $\si{\micro\second}$ and 3.1 $\si{\micro\second}$ for 
folding and unfolding respectively. Indeed, coarsening our MSM from nine to two macrostates gives us near perfect agreement with a folding 
MFPT of 14.0 $\si{\micro\second}$ and an unfolding MFPT of 3.0  $\si{\micro\second}$. The high temporal resolution models produced by the rapid convergence of our implied timescales with lagtime is likely the key reason for the high accuracy MFPT estimates from our SRV-MSM. Suarez et al.~\cite{Suarez2016}\ analyzed this same data using higher-order Markov approaches to report 
folding and unfolding MFPTs of 8.4 $\si{\micro\second}$ and 1.9 $\si{\micro\second}$, respectively. The discrepancy may be due to different macrostate definitions. 

\begin{table}[]
\begin{tabular}{crcr}
\textrm{transition} & \textrm{mean / \si{\micro\second}} && \textrm{std / \si{\micro\second}} \\
\hline
$\mathcal{S}_7 \to \mathcal{S}_{(0,1,2,3,4,5,6,8)}$ & 2.9  & $\pm$ & 0.2 \\
$\mathcal{S}_{(0,1,2,3,4,5,6,8)} \to \mathcal{S}_7$ & 16.3 & $\pm$ & 0.9
\end{tabular}
\caption{Calculated MFPTs into and out  of the folded state \state{7}.}
\label{table:mfpt}
\end{table}

\begin{figure*}
	\begin{center}
        \includegraphics[width=\textwidth]{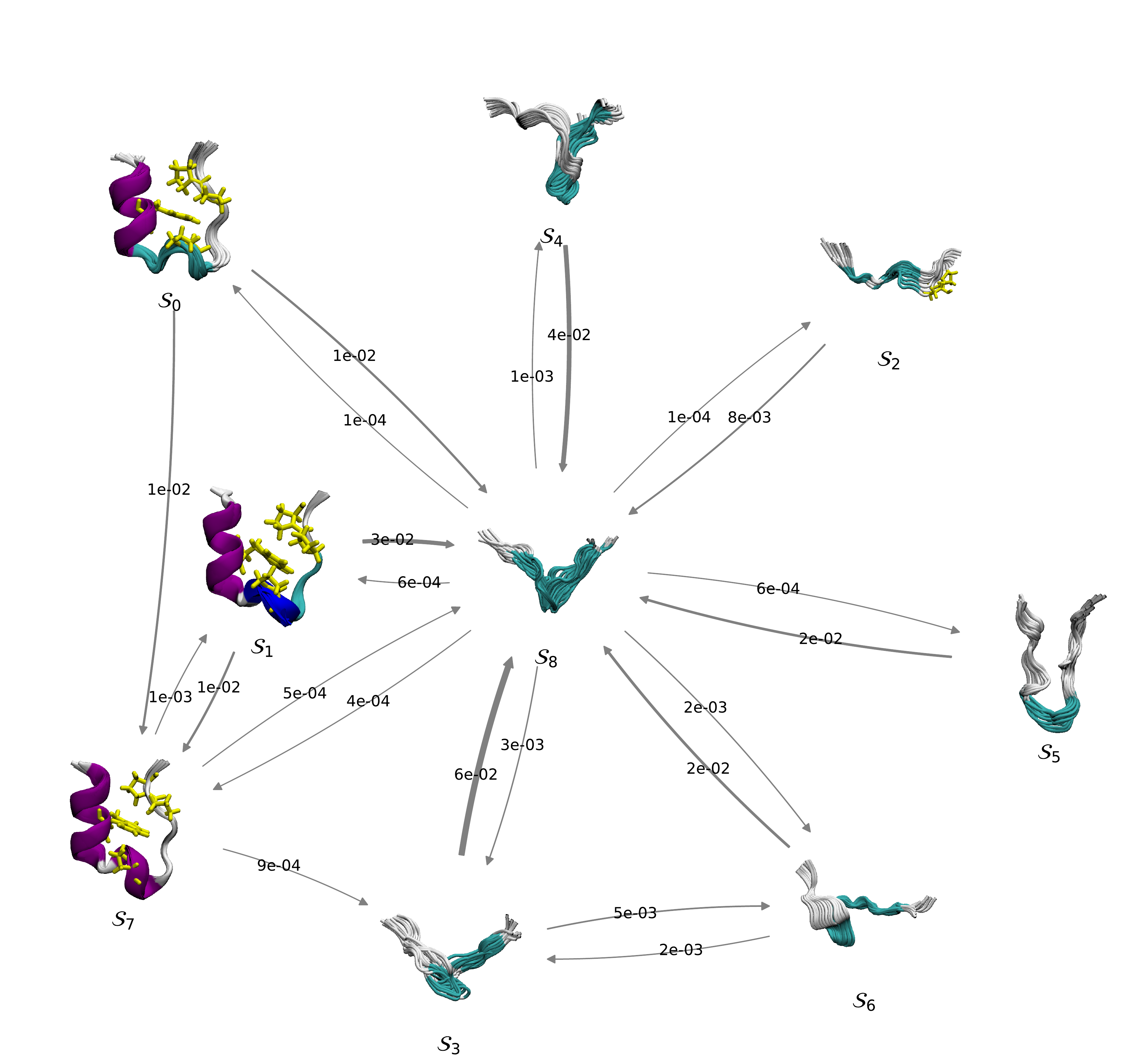}
        \caption{Visualization of the SRV-MSM for Trp-cage. The nine metastable macrostates \state{0-8} defined by PCCA\texttt{++} spectral clustering are represented by visualizations of an ensemble of 20 mutually aligned representative molecular states. The equilibrium transitions between states are represented by arrows and annotated by the transition probability corresponding to the associated off-diagonal macrostate transition matrix element. Arrow widths are drawn proportional to probabilities, and arrows corresponding to fluxes smaller than $10^{-5}$ are not visualized for clarity. The unfolded ensemble comprises \state{2,3,4,5,6,8} and is characterized by a central molten globule $\mathcal{S}_8$ that rapidly interconverts with the other metastable unfolded states. Folding into the native state \state{7} proceeds  either directly from \state{8}, through an intermediate \state{0} in which the $3_{10}$ loop is misfolded but stabilized by the hydrophobic core, or second intermediate \state{1} possessing an unfolded $3_{10}$ loop.
        }
        \label{fig:flux_viz}
	\end{center}
\end{figure*}

A visualization of the metastable conformational ensembles and the associated transitions of the nine macrostate SRV-MSM is presented in Figure~\ref{fig:flux_viz}. The stationary probabilities $\pi_{\mathcal{S}_i}$ and associated free energies $G_{\mathcal{S}_i}$ of each state \state{0-8} are listed in Table~\ref{table:equil}. The native fold resides in \state{7} and occupies $\sim$17\% of the stationary probability distribution at the $T$ = 290 K state point at which the molecular dynamics simulation was conducted.

\begin{table}[]
\begin{tabular}{ccc}
$\textrm{macrostate } \mathcal{S}_i$ & $\pi_{\mathcal{S}_i}$ & $G_{\mathcal{S}_i} / \textrm{k}_\textrm{B} T$ \\
\hline
$\mathcal{S}_0$ & 0.004837 & 5.332  \\
$\mathcal{S}_1$ & 0.008090 & 4.817  \\
$\mathcal{S}_2$ & 0.006681 & 5.009  \\
$\mathcal{S}_3$ & 0.016846 & 4.084  \\
$\mathcal{S}_4$ & 0.012673 & 4.368  \\
$\mathcal{S}_5$ & 0.020058 & 3.909  \\
$\mathcal{S}_6$ & 0.075622 & 2.582  \\
$\mathcal{S}_7$ & 0.168266 & 1.782  \\
$\mathcal{S}_8$ & 0.686928 & 0.376  \\
\end{tabular}
\caption{Stationary probabilities $\pi_{\mathcal{S}_i}$ and associated free energies $G_{\mathcal{S}_i}$ of each state \state{0-8} within the nine macrostate SRV-MSM.}
\label{table:equil}
\end{table}

The unfolded ensemble comprises \state{2,3,4,5,6,8} and accounts for $\sim$82\% of the stationary probability distribution. This ensemble is dominated by a central molten globule \state{8} that itself accounts for $\sim$69\% of the stationary distribution and acts as a kinetic hub for interconversions with the other unfolded metastable conformations. Of the remaining unfolded states, state \state{2} is particularly interesting and structurally interpretable. Transitions from \state{8} to \state{2} correspond to transitions along TIC1 in the free energy surface visualization in Figure \ref{fig:hde_msm_fes}a,f). Structurally, this transition can be identified as the rearrangement of the polyproline II structure (residues 17-20) from an unstructured to an alpha
helix-like conformation. In particular, transitions into \state{2} are defined by conversions of the Pro18 residue dihedrals from a $P_{||}$ ($\phi$ = -75$^\circ$, $\psi$ = 160$^\circ$) to an $\alpha$  ($\phi$ = -75$^\circ$, $\psi$ = -50$^\circ$) configuration. The native fold in \state{7} is stabilized by hydrophobic interactions
of the Trp6 side chain with Pro12, Pro18, and Pro19~\cite{Haabis2012}, and transitioning into \state{2} prohibits folding because the Pro18 rotates externally, facing away from the hydrophobic core and precluding stacking against the Trp6 side chain. Indeed, Figure~\ref{fig:hde_msm_fes}a,f shows the absence of any pathway along TIC1 from \state{2} to \state{7} and Figure~\ref{fig:flux_viz} shows the absence of any significant flux between these states. Instead, in order to fold the conformations in \state{2} must first transition into \state{8}, 
which effectively ``unlocks'' the molecule by enabling Trp6-Pro12 hydrophobic stacking.


The remaining unfolded macrostates, \state{3}, \state{4}, \state{5}, and \state{6}, together account for $\sim$13\% of the stationary probability distribution, and show negligible flux between one another or to the folded state \state{7} or intermediates \state{0} or \state{1}. Accordingly, folding is mediated through the compact molten globule \state{8} as evinced by the fact that the slowest timescale is associated with transitions from the unfolded to folded ensembles. In other words, mixing of the different unfolded states occurs at faster timescales than folding transitions, the flux of which is gated  almost exclusively through \state{8}. Deng et al.~\cite{Deng2013} indicate in their analysis of this simulation data that they find no evidence of kinetic partitioning of the unfolded state space, which is consistent with a hub-like scenario. Our observation of folding mediate by the molten globule state is also similar to an observation made by Marinelli et al.~\cite{Marinelli2009} in a study of the Trp-cage TC5b mutant, although they note a lower occupancy probability of the molten globule state. An analysis of the same D.E. Shaw trajectory as studied herein by Dickson and Brooks~\cite{Dickson2013} is also consistent with our model. They calculate a 
``hub score'' for the native state, defining the degree to which it mediates non-native-to-non-native transitions to determine that a substantial number of these transitions are not mediated by the native state.  

\begin{figure}
	\begin{center}
        \includegraphics[width=0.5\textwidth]{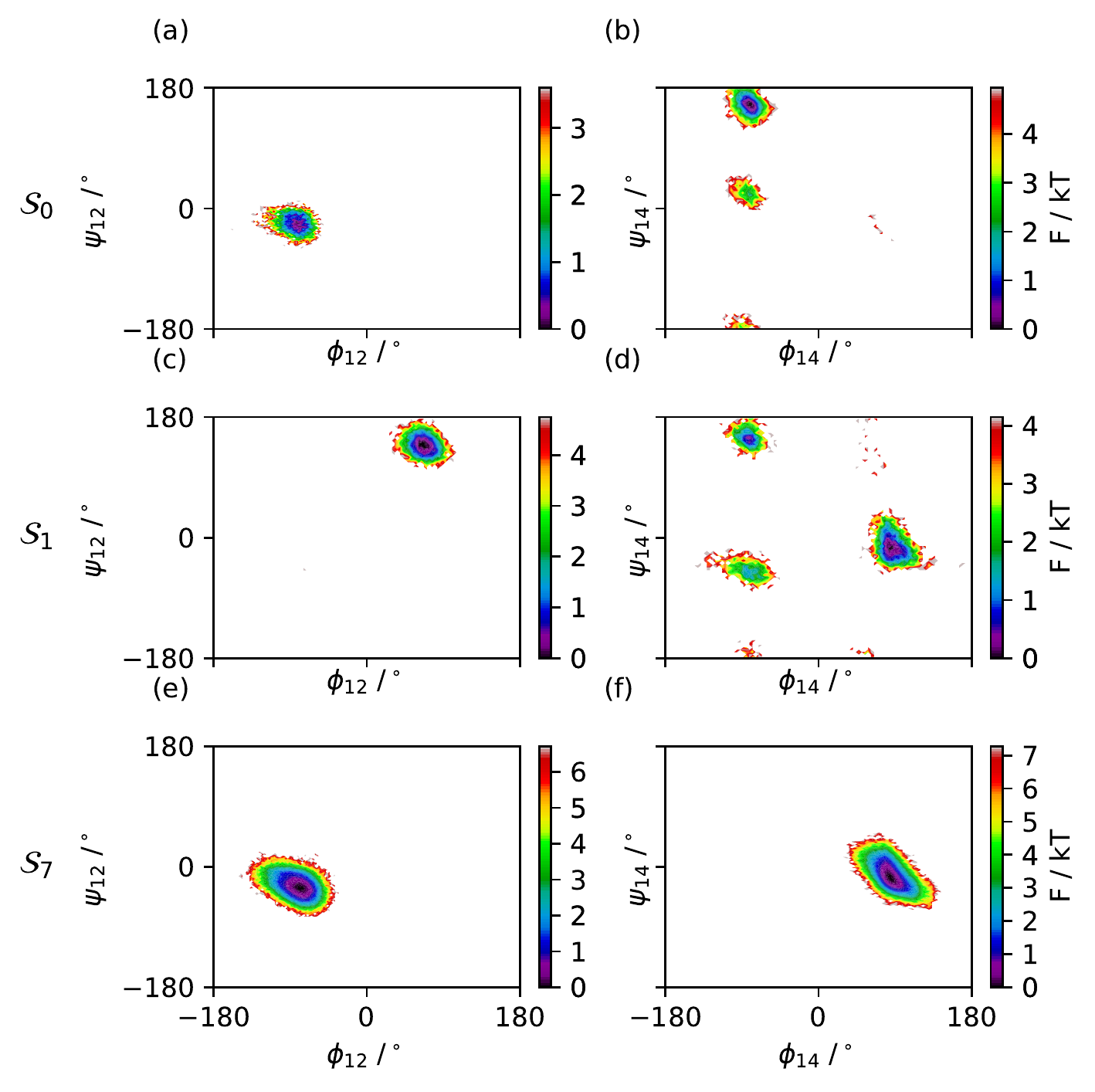}
        \caption{Ramachandran plots of backbone dihedrals of residues  Pro12 and Ser14 for states \state{0} (a-b), \state{1} (c-d), and \state{7} (e-f). The recognized ``loop structure'' \state{1} is distinguished by an unfolded $3_{10}$ loop as reflected in $\phi_{12}$-$\psi_{12}$ (c). The folding of the $3_{10}$ loop in \state{0} is disrupted by $\beta$ or $P_{||}$ character present in  $\phi_{14}$-$\psi_{14}$ (b) as opposed to the native $\alpha_L$ conformation (f). State \state{1} occupies both the $\beta$ or $P_{||}$ and $\alpha_L$ conformations in addition to $\alpha_R$ owing to the greater degree of flexibility of the  $3_{10}$ loop in this state.
        }
        \label{fig:folded_torsions}
	\end{center}
\end{figure}

\begin{figure}
	\begin{center}
        \includegraphics[width=\figurewidth]{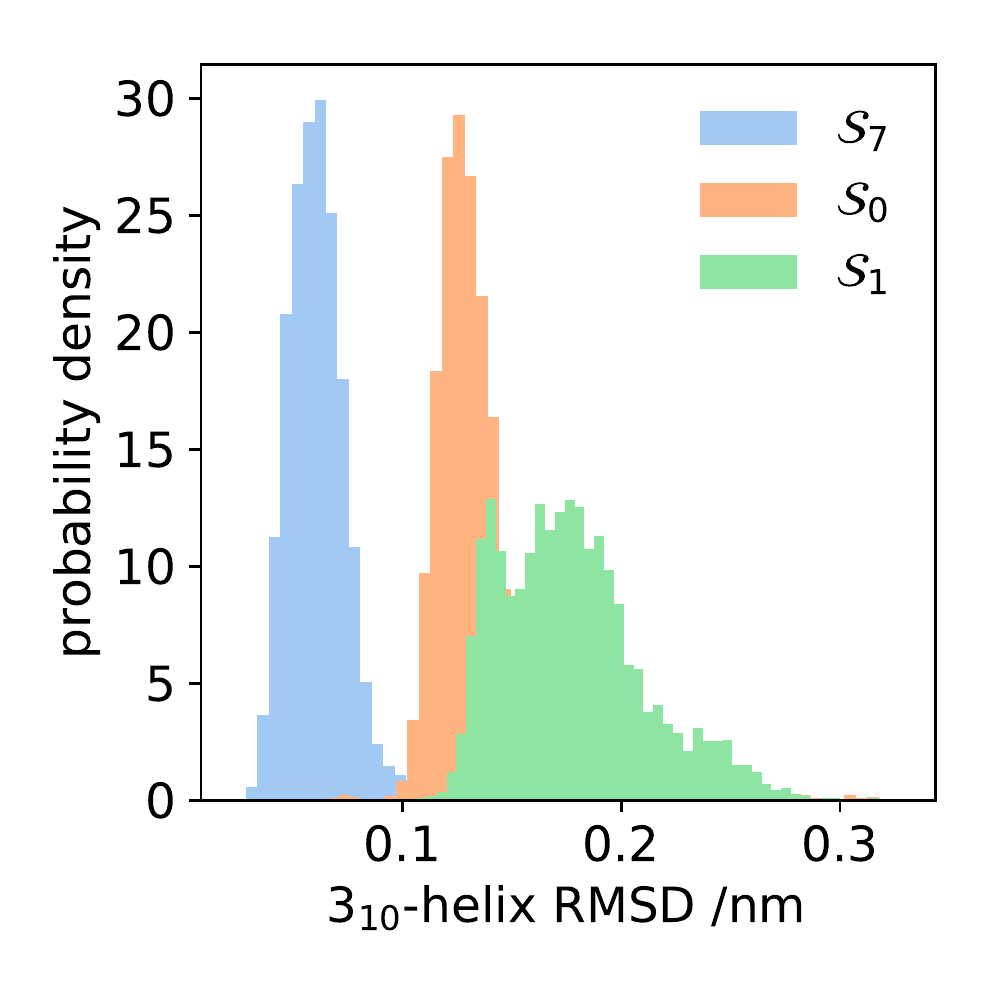}
        \caption{Distribution of the root mean squared deviation (RMSD) relative to the  native fold of residues 11-14 comprising the $3_{10}$-helix for states \state{0}, \state{1} and \state{5}. The folded state \state{7} and intermediate state \state{0} are both normally distributed with means of 0.065 nm and 0.135 nm, respectively. The intermediate \state{1} possesses a much broader non-normal distribution with mean 0.176 nm. 
        }
        \label{fig:310_dist}
	\end{center}
\end{figure}

Our model predicts folding to the native state  to proceed either directly from the molten globule kinetic hub $\mathcal{S}_8 \rightarrow \mathcal{S}_7$, or via one of two intermediates: 
$\mathcal{S}_8 \rightarrow \mathcal{S}_0 \rightarrow \mathcal{S}_7$, or $\mathcal{S}_8 \rightarrow \mathcal{S}_1 \rightarrow \mathcal{S}_7$. The intermediates \state{0} and \state{1} respectively occupy $\sim$0.48\% and $\sim$0.81\%  of the stationary probability  distribution, and bear a great deal of resemblance to both one another and to the native folded state. They are differentiated almost exclusively by the degree of folding of of the $3_{10}$-helix (residues 11-14).
Figure~\ref{fig:310_dist} shows the distributions of the root mean squared deviation (RMSD) from the native fold of these four $3_{10}$-helix residues for the simulation snapshots populating states \state{0}, \state{1} and \state{7}. The distributions for \state{0} and \state{7} are narrow and normal, indicating locally 
stable conformations. \state{1}, on the other hand, displays a much broader non-normal distribution. This may be characteristic of multiple states grouped together which cannot be separated at the temporal resolution of our model, or alternatively of a greater degree of flexibility in the motion of the $3_{10}$-helix region due to it being unfolded in this conformation. Focusing on the dihedral angles within the $3_{10}$-helix, Figure~\ref{fig:folded_torsions} displays the Ramachandran plots for residues 12 (a-c) and 14 (d-f). The unlooping of intermediate \state{1} relative to the native fold \state{7} can be seen most obviously in Pro12, with this residue transitioning from a native $\alpha$ ($\phi$ = -75$^\circ$, $\psi$ = -30$^\circ$) configuration to a $\alpha^{''}$ ($\phi$ = 75$^\circ$, $\psi$ = 145$^\circ$) configuration. The distinction between intermediate \state{0} and native state \state{7} is characterized by largely Ser14 $\beta$ character ($\phi$ = -80$^\circ$, $\psi$ = 155$^\circ$) in the former, compared to predominantly
$\alpha_L$ character ($\phi$ = 90$^\circ$, $\psi$ = -10$^\circ$) in the latter. The Ser14 residue in \state{1} shows occupancy $\beta$, $\alpha_L$, and $\alpha_R$ ($\phi$ = -80$^\circ$, $\psi$ = -20$^\circ$) configurations owing to the greater flexibility of the  $3_{10}$ loop in this state.

The state \state{1} is a well-known structural metastable state -- sometimes referred to as the ``loop'' structure -- in close proximity to the native fold but possessing an unfolded $3_{10}$-helix~\cite{Juraszek2006, Zhou2003, Wang2018, Kim2015}. This state \state{1} can be identified as a local minimum in Figure~\ref{fig:hde_msm_fes}b,g existing as a finger protruding below the  direct  path linking \state{8} and \state{7} along TIC1. Long-range interactions between the Trp6 core and Pro12 on the $3_{10}$ loop stabilize the \state{1} intermediate, and there is significant flux both into the native state \state{7} or back to the molten globule \state{8}. The state we identify as \state{0} does not receive much mention in the literature, likely due to its relative instability, possessing about half the stationary probability distribution compared to \state{1} (cf. Table~\ref{table:equil}) and about one sixth of the flux from the molten globule \state{8} (cf. Figure~\ref{fig:flux_viz}). In sum, folding proceeds from the molten globule hub \state{8} into \state{7,0,1} through the formation of the hydrophobic core in which the Trp6  sidechain is ``caged'' by the  Tyr3, Leu7, Gly11, Pro12, Pro18, and Pro19 sidechains, and the N-terminal $\alpha$-helix (residues 2-8). These structural formation events are either accompanied by complete folding of the $3_{10}$-helix (residues 11-14) through a direct transition into the native state \state{7}, or partial folding of the $3_{10}$-helix that leads to one or other of the metastable intermediates \state{0} or \state{1} that require subsequent structural rearrangements of the $3_{10}$ region to reach the native fold.

In summary, we have demonstrated the use of SRVs to establish a high time resolution SRV-MSM for the K8A mutant of Trp-cage TC10b at 290 K \cite{Lindorff-Larsen2011}. We carefully selected the model hyperparameters and verified its dynamic validity through cross-validation, spectral analysis, implied timescale convergence, and the Chapman-Kolmogorov test to present free energy surfaces and a macrostate transition model that sheds new understanding on its folding. In particular, we identify an unfolded ensemble dominated by a hub-like molten globule that mediates transitions to the folded state. Folding proceeds either directly through the simultaneous formation of the hydrophobic core, N-terminal $\alpha$-helix, and $3_{10}$-helix, or indirectly through one of two metastable intermediates that possess misfolded  $3_{10}$-helices. We note that our results differ from, although not necessarily inconsistent with, the folding model extracted from this data by Deng et al.~\cite{Deng2013} employing an RMSD-based MSM. Based on that analysis, Trp-cage folding was reported to proceed by two representative parallel paths as proposed by Juraszek and Bolhuis \cite{Juraszek2006} corresponding to two archetypal mechanisms of protein folding \cite{Kim2015,karplus1976protein,abkevich1994specific,gianni2003unifying}: (i) a nucleation-condensation mechanism wherein formation of a compact molten globule precedes folding of the N-terminal $\alpha$-helix, $3_{10}$-helix, and native packing of the hydrophobic core, and (ii) a diffusion-collision mechanism wherein pre-formation of the $\alpha$-helix precedes formation of the hydrophobic core and $3_{10}$-helix. The high resolution SRV-MSM established in this work establishes the dominance of a molten globule kinetic hub state that mediates folding. However, this statistical portrait is limited to the resolution of the 10 ns lagtime, and the fastest implied timescale resulting from PCCA++ macrostate clustering is $\sim$100 ns (Figure~\ref{fig:msm_its_cktest}a). 

Notwithstanding, our results present three important adjustments to the picture of the conformational and kinetic landscape. First, the 
presentation of two independent folding pathways, starting either from a molten globule or an extended conformation with a pre-formed helix can be limiting. The molten globule conformation serves as the gateway for folding and, within the statistical resolution supported by the data and our 
model, acts as a source for both folding pathways. Second, the unfolded ensemble possesses structural and kinetic richness centered upon this molten globule kinetic hub. Third, folding proceeds either directly to the native state, or through two non-native folded intermediates, which differ in the nativeness of the $3_{10}$-helix. The high kinetic resolution MSM enabled by the replacement of TICA by SRVs reveals a new intermediate \state{0} as an important metastable intermediate for folding. We note that it is not possible to resolve further structural details of the folding process by introducing additional macrostates into the SRV-MSM since analysis of the microstate transition matrix eigenvalue spectrum shows the simulation data to support no more than nine statistically robust macrostates. Resolution of finer-scale folding mechanisms and pathways from \state{8} to \state{7,0,1} would require a more detailed analysis 
of the microstate transition matrix, as previously studied by 
Deng et al.~\cite{Deng2013}, and/or path sampling calculations, which are beyond the scope of this work.

\section{Conclusions}

We have presented SRVs as viable and promising modular replacement for TICA in the construction of Markov state models for protein folding. In an application to an ultra-long 208 $\mu s$ explicit solvent simulation of the K8A mutant of Trp-cage TC10b conducted by D.E.~Shaw Research.~\cite{Lindorff-Larsen2011}, we showed SRV coordinates to outperform TICA-MSMs under cross validation by displaying higher test VAMP-2 scores with lower variance, and also to be more robust to input feature choices than TICA due to their capacity to learn nonlinear transformations of the 
input features. Employing SRVs as a basis set MSM 
construction produced a superior convergence rate of implied timescales with respect to lagtime, enabling the construction of extremely high resolution kinetic models. The resulting SRV-MSM revealed new understanding and insight into the kinetics and mechanisms of Trp-cage folding. A compact molten globular state acts as a kinetic hub for the unfolded ensemble and serves as the gateway for 
transitions into the folded state. The dominant folding pathway proceeds by formation of the hydrophobic core and N-terminal $\alpha$-helix either directly into native state or via one of two intermediates that possess imperfectly folded $3_{10}$-helices. The high time resolution MSMs enabled by SRVs represent a valuable new addition to the MSM construction pipeline that can help squeeze the most out of the simulation data used to parameterize the models and produce high-temporal resolution kinetic models to understand and predict biomolecular folding.


\section*{Acknowledgments}
	
	This material is based upon work supported by the National Science Foundation under Grant No.~CHE-1841805.
	H.S. acknowledges support from the Molecular Software Sciences Institute (MolSSI) Software Fellows program 
	(NSF grant ACI-1547580) \cite{krylov2018perspective,wilkins2018nsf}. We are grateful to D.E. Shaw Research for sharing the Trp-cage simulation trajectories.

\bibliography{references}
\bibliographystyle{apsrev4-1}

\end{document}